\documentstyle[11pt,paspconf]{article}

\begin{document}

\title{The Identification of Ly$\alpha$ Absorbers at Low Redshift}
\author{Suzanne M. Linder}
\affil{Pennsylvania State University, University Park, PA 16802 USA; 
slinder@astro.psu.edu}


Ly$\alpha$ absorption lines are useful for probing the properties of
gas in and around galaxies.
While it is known that some stronger
absorption lines arise in galaxies, the nature of 
weaker absorbers is more difficult to establish.  It is
unknown how weak absorption lines can be which are still associated
with galaxies.  One
common way to test for a relationship between absorbers and galaxies is to
plot equivalent width or neutral hydrogen
column density versus impact parameter between the observed galaxies and the
quasar lines of sight.
Such plots have been used to support a variety of conclusions
about the relationship between galaxies and weak Ly$\alpha$ absorbers.
Chen et al. (1998) observed an anticorrelation which they argue supports
absorbers arising in 160 kpc halos around luminous, high surface brightness galaxies.
Bowen, Blades \& Pettini (1996) observed no such anticorrelation and thus argued
that absorbers often do not arise in ordinary galaxies.  Simulations by
Linder (1998a) show that if absorption is produced by galaxies with a wide
range of properties, an anticorrelation will occur, but a large amount of
scatter will be seen due to variations in galaxy properties.  In other studies
where an anticorrelation is seen, it is argued that absorption arises from
large scale structure rather than gas associated with particular galaxies
(Tripp, Lu \& Savage 1998; Dav\'e et al. 1998).

I 'observe' my simulated galaxies again (as in Linder 1998b) except using 
$M_B<-18$.  It is shown in the figures that an anticorrelation between
impact parameter and column density can occur largely as a result of selection
effects.  The nearest galaxy is likely to be observed at an impact parameter around
a few hundred kpc whether it gives rise to absorption or not, unless a
strong absorption line is seen.  Thus while it is clear that at least some
stronger ($>10^{16}$ cm$^{-2}$) absorption lines arise in galaxies, making
such plots does not allow for a meaningful test
of the relationship between galaxies and weaker absorbers.  While there
is no reason to think that gas around galaxies should be cut off at any
particular column density, other tests, such
as those in which the absorption cross sections of galaxies are examined
for a wide range of galaxy properties, will be needed to establish the 
relationship between absorbers and galaxies.

\begin{figure}[hb]
\plotfiddle{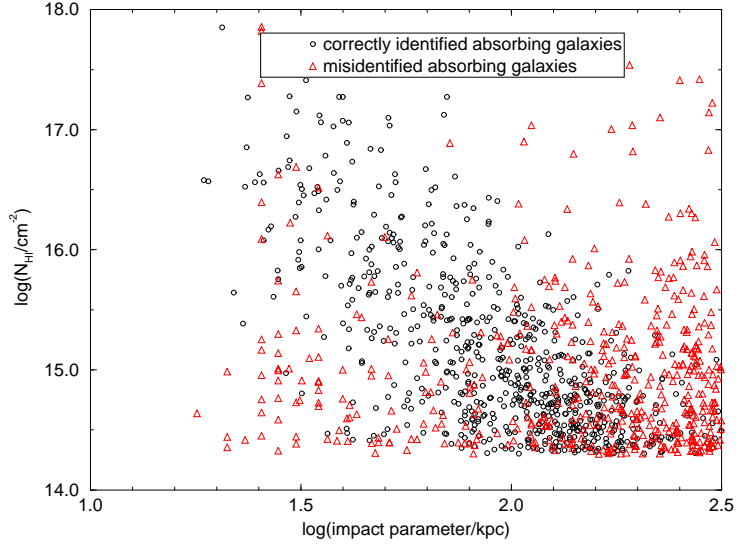}{2.5in}{270}{45}{45}{-164}{255}
\vskip  -0.25in
\caption{Neutral hydrogen column densities are plotted versus impact parameter
(between the galaxy and quasar line of sight)
for simulated absorbing galaxies which are 'observed' using the selection 
criteria defined in the text.  Circles show absorbing galaxies which are
identified correctly as the actual absorbing galaxies from the simulations.
Triangles show galaxies which are misidentified by the 'observations.'
The actual absorbing galaxy is frequently not identified for both weak
and strong Ly$\alpha$ absorbers, but it is less obvious to an observer
when a weak line is misidentified because the actual absorbing galaxy
is likely to be at a similar impact parameter.}
\label{fig1}
\end{figure}
\begin{figure}[bt]
\plotfiddle{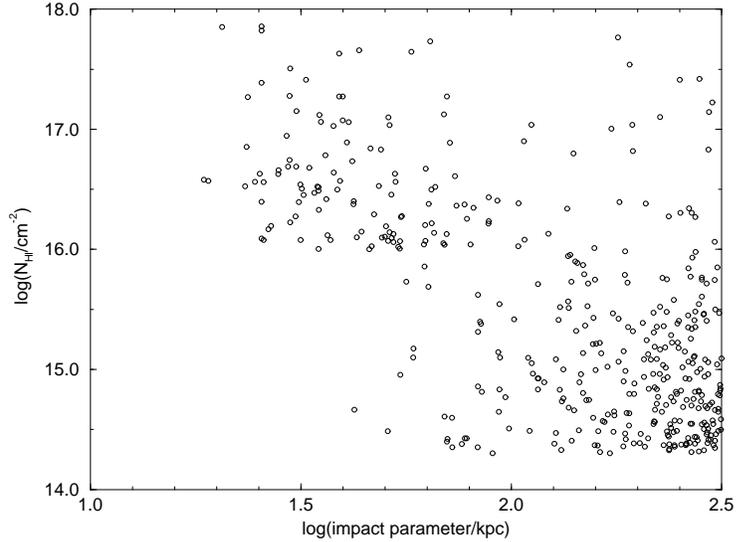}{2.5in}{270}{45}{45}{-164}{255}
\vskip  -0.25in
\caption{Neutral column densities are plotted versus galaxy impact parameter
for simulated, absorbing ($>10^{16}$ cm$^{-2}$) galaxies and
randomly distributed weaker absorbers, where the galaxies are 'observed' as 
above.
An anticorrelation can also arise if weaker absorbers are not associated
with galaxies in any way.}
\label{fig2}
\end{figure}


\begin{references}
\reference Bowen, D. V., Blades, J. C., \& Pettini, 1996, \apj, 464, 141
\reference Chen, H. -W., Lanzetta, K. M., Webb, J. K., \& Barcons, X. 1998, \apj, 498, 77
\reference Dav\'e, R., Hernquist, L., Katz, N., \& Weinberg, D. H. 1998, \apj, submitted
\reference Linder, S. M. 1998a, \apj, 495, 637
\reference Linder, S. M. 1998b, 171st IAU Proceedings (in press, astro-ph/9810162)
\reference Tripp, T. M., Lu, L., \& Savage, B. D. 1998, \apj, 508, in press
\reference 
\end{references}
\end{document}